\documentclass[twocolumn,showpacs,preprintnumbers,amsmath,amssymb]{revtex4}

\usepackage[T1]{fontenc}
\usepackage{graphicx}
\usepackage{dcolumn}
\usepackage{bm}

\begin{document}

\title{On the Density Dependent Nuclear Matter Compressibility}

\author{V.A. Dexheimer}
 \email{dexheimer@th.physik.uni-frankfurt.de}
\affiliation{FIAS, Johann Wolfgang Goethe University, Frankfurt am Main, Germany}

\author{C.A.Z. Vasconcellos, B.E.J. Bodmann}
\affiliation{Universidade Federal do Rio Grande do Sul, Porto Alegre, RS, Brazil}

\date{\today}

\begin{abstract}
In the present work we apply a quantum hadrodynamic effective model in the mean-field approximation to the description of
neutron stars. We consider an adjustable derivative-coupling model and study the parameter influence on the dynamics of the system by analyzing the full range of values they can take. We establish a set of parameters which define a specific model that is able to describe phenomenological properties such as the effective nucleon mass at saturation as well as global static properties of neutron stars (mass and radius). If one uses observational data to fix the maximum mass for neutron stars by a specific model, we are able to predict the compression modulus of nuclear matter $K = 257,2MeV$.
\end{abstract}

\maketitle

\section{Introduction}

Effective theories that implement hadronic dynamics are considered the most efficient way to describe infinite nuclear matter because they allow to work in a specific energy scale while ignoring other degrees of freedom of the system. In the present work nuclear matter is considered at low temperature and low chemical potential, thus the relevant degrees of freedom are hadrons. Typically two pathways are explored to get insight in hadronic matter properties, on the one side at high temperatures and low densities, heavy ion collisions, on the other hand exploring neutron stars at low temperatures and extreme densities \cite{Yagi2006}. With the increase of temperature in the first case, or density in the second, the quarks deconfine and chiral symmetry is effectively restored. Although this feature can be relevant for the study of neutron star matter, the existing models able to describe it \cite{chiral1,chiral2,han} are very complex and do not allow a straightforward analysis of more complex functional dependence of the couplings, that is exactly the intention of this work. In this article we present a generalized model theory for hadron dynamics in a nuclear matter environment, supposed to exist in neutron stars, and discuss the density dependent nuclear matter compressibility and its response to the parameters of the theory. In an initial step the model theory is calibrated for densities where from the experimental point of view phenomenology is well under control. In a second step, the undetermined parameters are partially fixed by a symmetry scheme or varied freely. The principal feature of our model theory is that it unites properties of a variety of approaches known in the literature (\cite{VADIWARA06} and references therein) at the cost of additional parameters. Nevertheless, as the following discussion will show, only a straight range for those parameters is physically meaningful, so that the increase of degrees of freedom is considerably reduced by phenomenology.

Further we make profit from the fact that small variations in the parameters create considerable differences in the results. For example a change in the binding energy of nuclear matter of $0.25 MeV$ implies a decrease by $0.5$ solar masses for the maximum mass as predicted by a model for a specific family of neutron stars as shown in more detail below. For that reason a precise knowledge of input parameters is crucial in order to make predictions relevant for observation. In the present work the input parameters are: the binding energy and the asymmetry coefficient of nuclear matter, the effective mass of the baryons and the compression modulus at the nuclear matter saturation point. The compression modulus may be understood as a measure for the difficulty to remove the system from equilibrium or for the affinity to return to equilibrium.

Until now, compressibility is one of the parameters determined with major uncertainty. One of the goals of the present work is to show the influence of the onset of hyperon presence and how this might be useful in pinning down the uncertainty in the compression modulus using observational evidences. To this end we calculate the compressibility for a large range of densities including the usually determined value at saturation density.

\section{The Model}

The basis for our calculations is a Quantum Hadrodynamics motivated model including beside the nucleons also the hyperons which define the baryon octet ($p$, $n$, $\Lambda$, $\Sigma^+$, $\Sigma^0$, $\Sigma^-$, $\Xi^0$ and $\Xi^-$). The nuclear medium effect is taken into account by a non-linear coupling involving the sigma field with adjustable parameters $\lambda$, $\beta$ and $\gamma$, which tune the intensity of the meson baryon couplings (for details see \cite{teseVAD2006,VADIWARA06}). The considered mesons are the usual scalar $\sigma$ and the isoscalar vector meson $\omega$ but also the isovector vector meson $\rho$. Additionally the lepton degrees of freedom ($e$ and $\mu$) establish or complement charge neutrality, depending on the density. The Lagrangian density reads:
\begin{eqnarray}\label{lag}
\mathcal{L}_T &=& \sum_{B}\bar{\psi}_B[\gamma_{\mu}(i\partial^{\mu}-g^*_{\omega B}\omega^{\mu}-\frac{1}{2}g^*_{\rho B}\mbox{\boldmath
$\tau.\rho^{\mu}$})
\nonumber \\
&&-(M_B-g^*_{\sigma B}\sigma) ]\psi_B + \sum_{\ell}\bar{\psi}_{\ell}(i\gamma_{\mu}\partial^{\mu}-m_{\ell})\psi_{\ell}
\nonumber \\
&&+\frac{1}{2}(\partial_{\mu}\sigma\partial^{\mu}\sigma-m_{\sigma}^2\sigma^2) - \frac{1}{4} \omega_{\mu\nu}\omega^{\mu\nu} +\frac{1}{2}
m_{\omega}^2\omega_{\mu}\omega^{\mu}
\nonumber \\
&& -\frac{1}{4}\mbox{\boldmath
$\rho_{\mu\nu}\rho^{\mu\nu}$}+\frac{1}{2}
m_{\rho}^2\mbox{\boldmath
$\rho_{\mu}\rho^{\mu}$},
\end{eqnarray}
where the index $B$ denotes a specific baryonic species and the index $\ell$ a specific leptonic species.
Here the adjustable coupling strengths are related with the usual coupling constants through
\begin{eqnarray}\label{cplng}
g^*_{\sigma B}&\equiv& \chi^*_{\lambda B}g_{\sigma B},
\nonumber \\
g^*_{\omega B}&\equiv& \chi^*_{\beta B}g_{\omega B},
\nonumber \\
g^*_{\rho B}&\equiv& \chi^*_{\gamma B}g_{\rho B},
\end{eqnarray}
and the adjustable coefficients $\chi^*_{i B}$ that drive the dynamical influence of the hadronic medium are
\begin{equation}\label{medf}
\chi^*_{i B}\equiv \left(1+\frac{g_{\sigma}\sigma
}{iM_B}\right)^{-i},
\end{equation}
for $i=\lambda$, $\beta$ or  $\gamma$ (see eq. \ref{cplng}). These parameters are constrained in order to reproduce phenomenology, in the present case an effective mass at saturation between $0.7$ and $0.8$ nucleon masses and a compression modulus between $200MeV$ and $300 MeV$. The coupling constants $g_{\sigma N}, g_{\omega N}$ and $g_{\rho N}$ of the nucleons are chosen in a way to have zero pressure and a determined energy at saturation (calculated assuming a binding energy of $-16 MeV$ and a density of $0.17 fm^{-3}$ at saturation and a bare nucleon mass of $939 MeV$) and to reproduce an asymmetry coefficient of $32.5 MeV$. Since no stringent experimental confirmation with respect to meson hyperon couplings is available and also possible nuclear medium effects on that couplings are an open question, one way to reduce these ambiguities is to use the $SU(6)$ flavor spin symmetry scheme as in ref. \cite{SU6} as a guide.
\begin{eqnarray}\label{gHyp}
\begin{array}{ccccc}
g_{\sigma \Sigma}=2/3g_{\sigma N} & \qquad & g_{\sigma \Xi}=1/3g_{\sigma N} & \qquad & g_{\sigma \Lambda}=2/3g_{\sigma N} \\
g_{\omega \Sigma}=2/3g_{\omega N} & \qquad & g_{\omega \Xi}=1/3g_{\omega N} & \qquad & g_{\omega \Lambda}=2/3g_{\omega N} \\
g_{\rho \Sigma}=2g_{\rho N} & \qquad & g_{\rho \Xi}=g_{\rho N} & \qquad & g_{\rho \Lambda}=0.
\end{array}
\end{eqnarray}

As already indicated before, one advantage of our model is that it is general in the sense, that certain choices of parameters $\lambda, \beta, \gamma$, permit to recover the principal models cited in the literature. For $\lambda=\beta=\gamma=0$ the Walecka model \cite{Walecka} is reproduced, for $\lambda =1$ and $\beta=\gamma=0$ the ZM1 model \cite{ZM1} is recovered, for $\lambda=\beta=\gamma=1$ the ZM3 model \cite{ZM1} results and $\lambda=\beta=\gamma\to\infty$ describes the exponential model \cite{ExpMod}. Beside the possibility to create models using intermediate values of the parameters, we can group the parameters according to the combinations in which they are being varied:
\begin{enumerate}
\item $\lambda$ varies with $\beta=\gamma= 0$, 
\item $\lambda$ varies with $\lambda=\beta=\gamma$,
\item $\lambda$, $\beta$ and $\gamma$ vary independently.
\end{enumerate}

For the first coupling scheme we find acceptable values for the effective mass and the compression modulus at saturation for  $0.06<\lambda<0.19$. For the coupling scheme 2 there are no acceptable values, which shows that the sigma field influence on the vector mesons is not ``universal''. For the last coupling scheme there are acceptable values for $0.00<\lambda<20.22$ and $0.00<\beta<1.35$, where both parameters correlate \cite{VADIWARA06,teseVAD2006}. A variety of combinations have shown us that only a considerable restricted choice of parameter combinations is possible in order to reproduce phenomenological meaningful quantities. The $\gamma$ parameter cannot be related to symmetric nuclear matter saturation properties as it relates to the $\rho$ meson and consequently to the isospin asymmetry.

Depending on the density, the constituent particles of the system change. The hyperons decay into other hyperons so that baryon number conservation and chemical equilibrium have to be taken into account. Note, that the chemical potential is the criterion which indicates the threshold where heavier baryons start to contribute to the chemical composition of nuclear matter. Further charge neutrality shall hold, according to findings in the literature, that a spherically symmetric charged neutron star is not stable \cite{Q0NS}. To solve the system analytically and in a self-consistent way the mean-field approximation is used, since for higher densities the fluctuations of the meson fields are negligible (see for instance ref. \cite{Walecka}).

\begin{figure}[!pt]
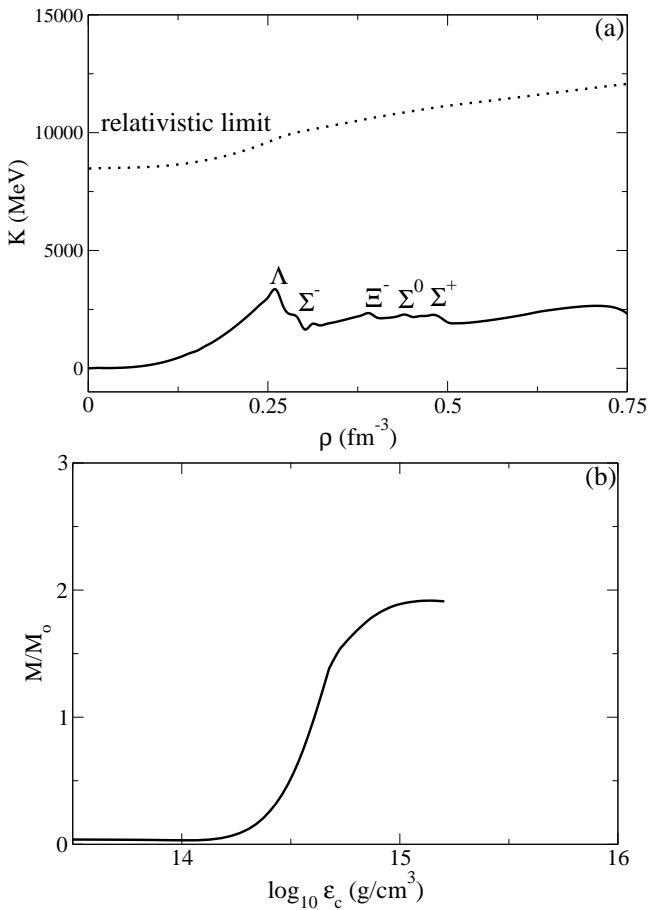

\begin{center}
\includegraphics[width=0.475\textwidth,clip,trim=0 0 0 0]{k0octeto.eps}
\includegraphics[width=0.45\textwidth,clip,trim=0 0 0 0]{mwalrhoocteto.eps}
\end{center}
\caption{Compressibility as a function of baryon density (a) and mass of the star as a function of central energy density (b) for $\lambda=0$ (Walecka model).}
\label{wal+}
\end{figure}

\begin{figure}[!pt]
\begin{center}
\includegraphics[width=0.35\textwidth,clip,trim=0 40 0 0]{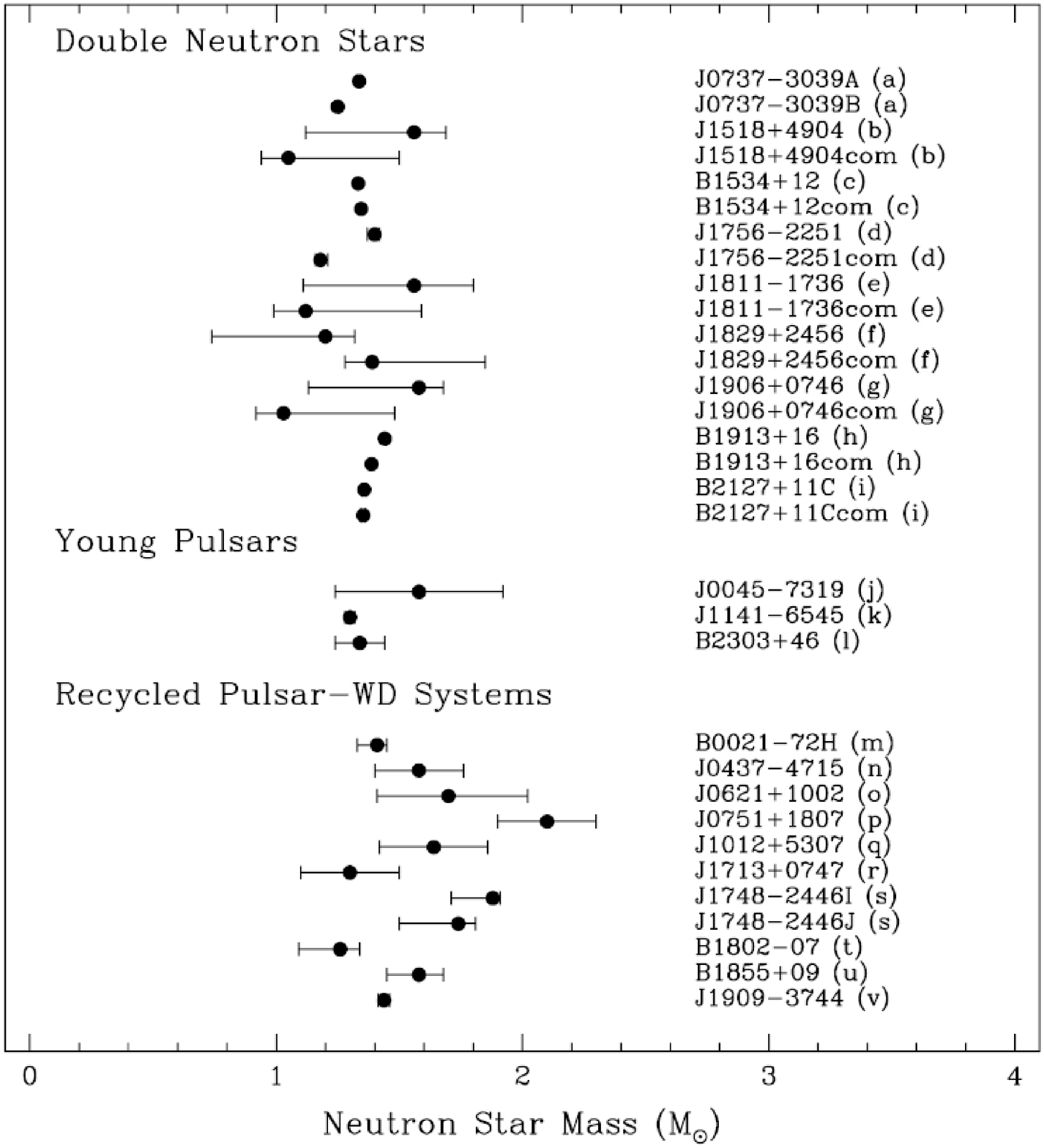}
\end{center}
\caption{Observed masses of known pulsars in units of solar masses.}
\label{NSMass}
\end{figure}

\section{Compressibility Function}

The compressibility function of the star as a function of density is determined from its thermodynamical definition (proportional to the derivative of pressure with respect to volume) \cite{teseVAD2006,VADIWARA06} and written as the first derivative of pressure $P$ (with respect to the total baryon density), the second derivative of the energy density $\epsilon$, or using the definition of baryon chemical potential $\mu_n=\partial\epsilon/\partial \rho$:
\begin{equation}\label{muB}
K=9 \frac{\partial P}{\partial \rho}=9\rho\frac{\partial ^2\epsilon}{\partial \rho}=9\rho\frac{\partial \mu_n}{\partial \rho} \; ,
\end{equation}
where $\mu_n$ is equal to the chemical potential for neutrons. The chemical potential for different baryon species can be calculated from:
\begin{equation}
\mu_B=\sqrt{k_{F_B}^2+(M_B^*)^2}+(g^*_{\omega
B})_0\omega_0+(g^*_{\rho B})_0\rho_{03}I_{3B} \; .
\end{equation}
For each density the chemical equilibrium equations determine the chemical potential for each species which determines its Fermi energy. Once the particle's chemical potential is less than the Fermi energy the system favors the presence of the respective particle. Here the chemical potential of each species is written as a function of the effective coupling strengths, and isospin is explicitly included. Isospin effects and charge conservation together with the nontrivial response of the effective mass to the presence of the hadronic environment manifest themselves in the phenomenon that hyperons composing the star matter as a function of density do not appear simply according to their vacuum mass hierarchy. Fig. \ref{wal+} (a) shows this behavior in the compressibility function as a function of baryon density. 

In agreement with relativistic formulations this model respects the causal limit ( i.e., the speed of sound in the medium is smaller than the speed of light for any density), so from the relation of compressibility and baryon chemical potential the speed of sound in the medium is
\begin{eqnarray}
\left(\frac{v}{c}\right)^2=\frac{K}{9\mu_n},
\end{eqnarray}
and the following inequality is established
\begin{eqnarray}
K<9\mu_n.
\end{eqnarray}
In Fig. \ref{wal+} (a) this fact is indicated by the limiting dotted line in the respective plot of the compressibility function.

\section{Neutron Stars}

\begin{figure}[!pt]
\begin{center}
\includegraphics[width=0.475\textwidth,clip,trim=0 0 0 0]{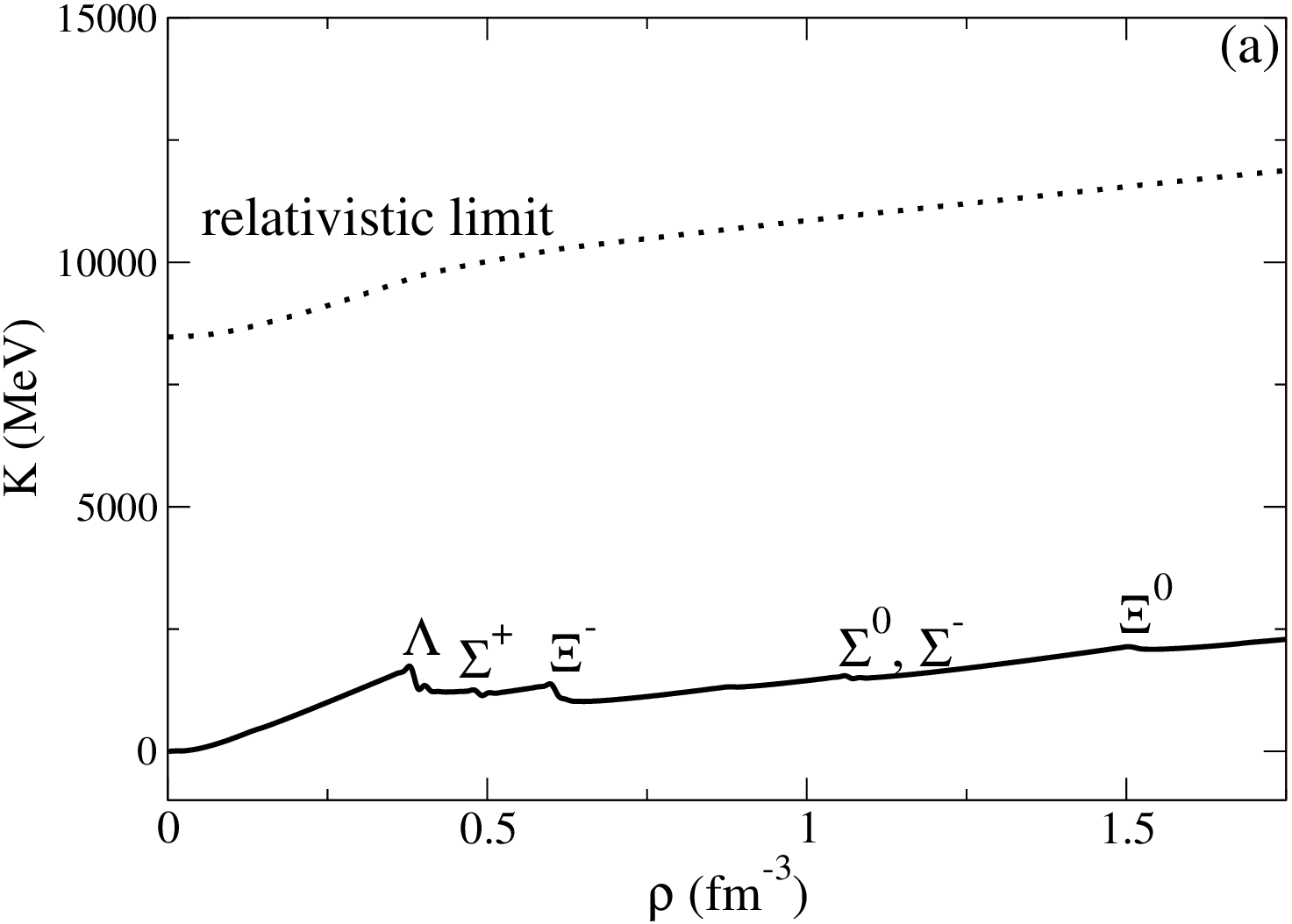}
\includegraphics[width=0.45\textwidth,clip,trim=0 0 0 0]{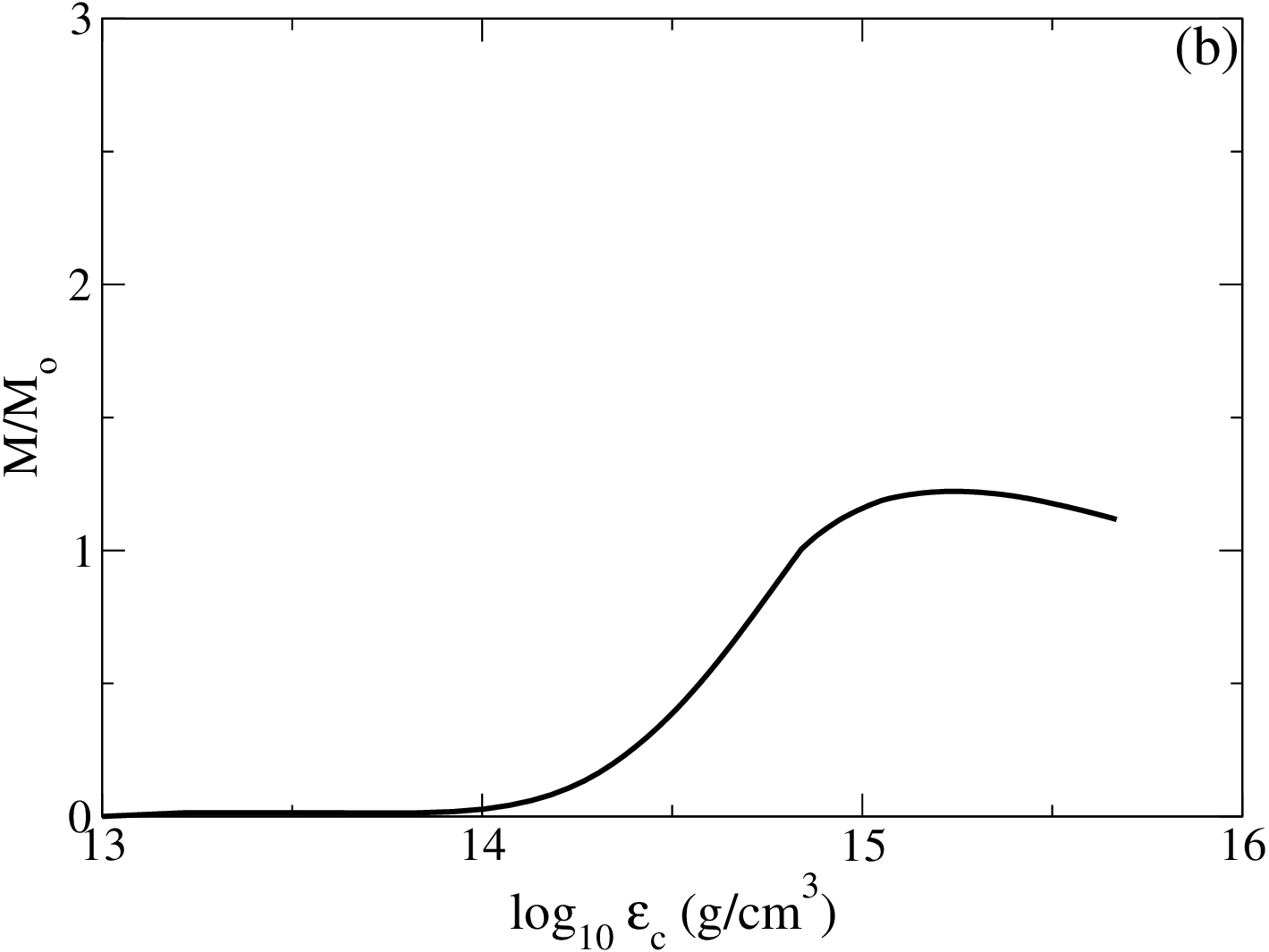}
\end{center}
\caption{Compressibility as a function of baryon density (a) and mass of the star as a function of central energy density (b) for $\lambda\to\infty$.}
\label{laminf}
\end{figure}

\begin{figure}[!pt]
\begin{center}
\includegraphics[width=0.475\textwidth,clip,trim=0 0 0 0]{Kexplambdabeta.eps}
\includegraphics[width=0.45\textwidth,clip,trim=0 0 0 0]{mexplambdabeta.eps}
\end{center}
\caption{Compressibility as a function of baryon density (a) and mass of the star as a function of central energy density (b) for $\lambda$, $\beta\to\infty$.}
\label{lambetinf}
\end{figure}

\begin{figure}[!pt]
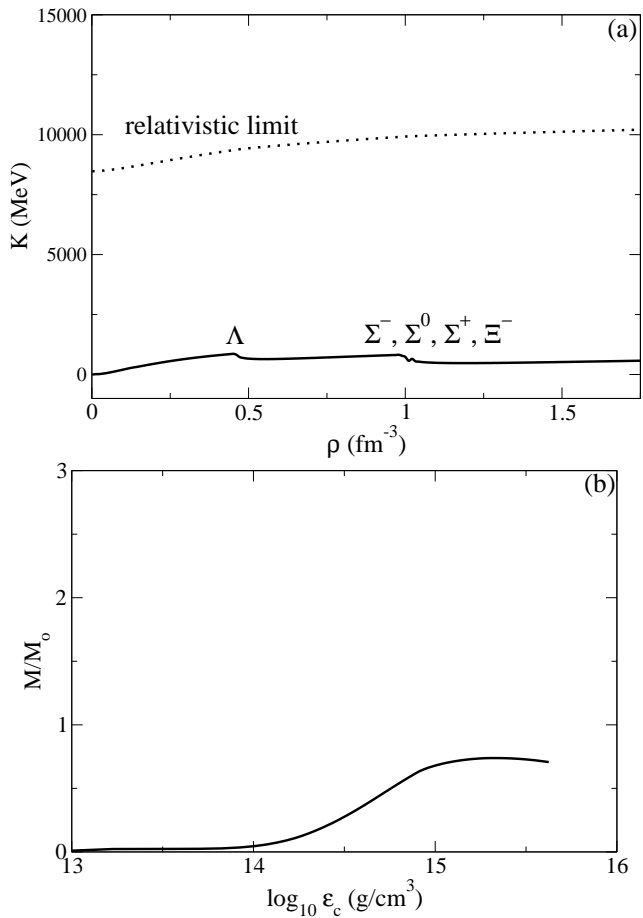

\begin{center}
\includegraphics[width=0.47\textwidth,clip,trim=0 0 0 0]{Kexp.eps}
\includegraphics[width=0.45\textwidth,clip,trim=0 0 0 0]{mexplambdabetagamma.eps}
\end{center}
\caption{Compressibility as a function of baryon density (a) and mass of the star as a function of central energy density (b) for $\lambda$, $\beta$, $\gamma\to\infty$ (exponential model).}
\label{lambetgaminf}
\end{figure}

\begin{figure}[!pt]
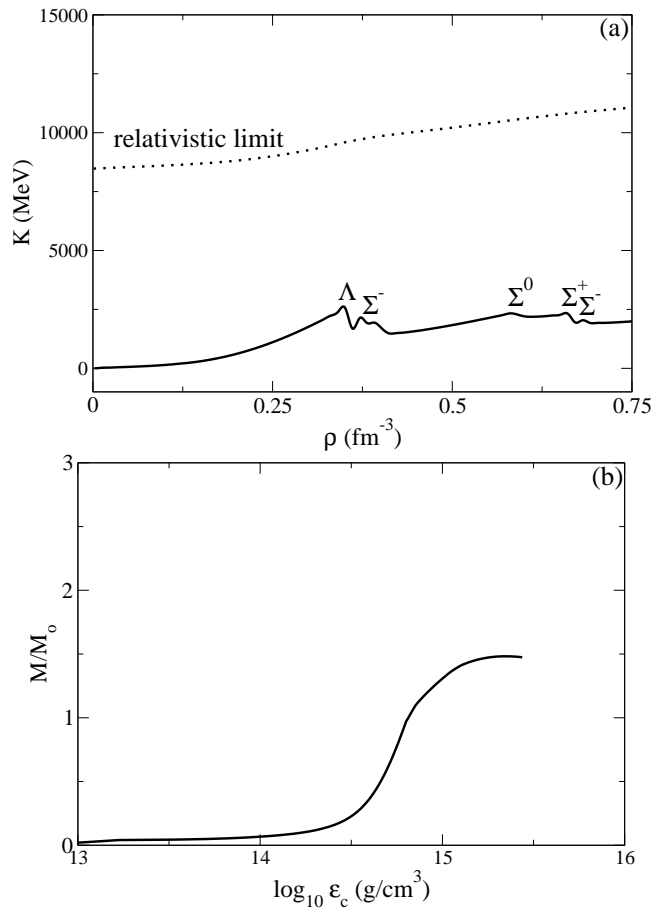

\begin{center}
\includegraphics[width=0.475\textwidth,clip,trim=0 0 0 0]{kl00631gm1.eps}
\includegraphics[width=0.45\textwidth,clip,trim=0 0 0 0]{ml006b0g-1.eps}
\end{center}
\caption{Compressibility as a function of baryon density (a) and mass of the star as a function of central energy density (b) for $\lambda=0.06$, $\beta=0$, $\gamma=-1$.}
\label{best}
\end{figure}

The equation of state obtained from the Lagrangian density (\ref{lag}) is then plugged into the Tolman-Oppenheimer-Volkoff equation, which adds gravitational effects to nuclear matter properties. For each of the parameter sets, a family of neutron stars is obtained and compared with observational data. Then it is possible to analyze the behavior of the $\gamma$ parameter in the neutron star mass and radius so that the best parameter set may be chosen.

As can be seen in the Fig. \ref{NSMass}, within the observational error bars the mass of almost all example pulsars have a value approximately $1.4 M_\odot$. However, lately pulsars with higher masses (the largest one with a mass of $2.1\pm0.2 M_\odot$) have been observed. Since there is no criterion as to which model (QHD, quark matter or others) is the most adequate starting point in order to derive the equation of state, we do not consider the largest possible mass as a relevant reference for the present model and focus on finding the model that agrees with nuclear phenomenology and predict the largest possible neutron star mass ($\sim 1.5 M_\odot$). The radius of the model stars range between $10$ km and $14$ km, which because of uncertainties in obtaining stringent values from observation, does not permit to put further constraints on the parameters.

The model with all the parameters equal to zero, which corresponds to the Walecka model extended by the inclusion of hyperons from the baryon octet and the meson $\rho$ predicts a larger maximum mass $1.9 M_\odot$ (Fig. \ref{wal+} (b)) but suffers a drawback because of a considerable small effective mass $M^* \sim 0.5 M$ and an amplified compression modulus $K \sim 500 MeV$.

An increase in $\lambda$, while keeping $\beta$ e $\gamma$ equal to zero, decreases the compressibility function (Fig. \ref{laminf} (a)) because the star can only support a lower mass against gravity (Fig. \ref{laminf} (b)). This behavior happens for values $\lambda<1.75$. After that, the system saturates, meaning that it does no respond to further changes in the parameter. The range with acceptable nuclear phenomenology is between $0.06<\lambda<0.19$ and corresponds to maximum mass from $1.46$ to $1.22$ solar masses.

Upon increasing $\lambda$ and $\beta$ together, keeping $\gamma$ equal to zero, the compressibility function (Fig. \ref{lambetinf} (a)) and the neutron star mass (Fig. \ref{lambetinf} (b)) decrease. A variety of combinations for $\lambda$ and $\beta$ (not shown in this article, but elsewhere \cite{teseVAD2006}) have shown us that $\beta$ and $\lambda$ have a similar effect on the macroscopic properties, but an increase in $\beta$ shifts the onset of saturation towards larger values for $\lambda$, which saturates for $\lambda>16$ and $\beta>10$. The range with acceptable nuclear phenomenology ranges between $0.00<\lambda<20.22$ and $0.00<\beta<1.35$.

As the third option a variation of $\gamma$, which tunes the isospin contribution in nuclear matter, also (slightly) modifies the compressibility function (Fig. \ref{lambetgaminf} (a)) and neutron star mass (Fig. \ref{lambetgaminf} (b)), respectively. Depending on the isospin sign of the particle (for example positive for proton and negative for neutron), the contribution of the term related to $\gamma$ can be positive or negative making the net effect small in comparison with $\lambda$ and $\beta$ contributions.

\section{Conclusion}

In the present article we calculated the density dependent compressibility function for a generalized model for nuclear matter. Usually, compressibility is determined only at saturation density, however, in order to characterize states in the final phase of a neutron star evolution, where quasi equilibrium conditions are encountered, a large density range for the compressibility gives details relevant for possible processes.

With increasing density, hyperons come into play and soften the equation of state. The couplings have shown to be of crucial influence in the conditions which determine the sequence in which hyperons appear. Because the present model mimics medium effects by the dependence of the couplings on the sigma field, this nonlinearity is responsible for the relevant hyperon hierarchy, which does not reflect the sequence of threshold by simply considering the masses.

Among the variety of possible models, the one in agreement with nuclear phenomenology and astrophysical observation is defined by the parameter set $\lambda=0.06$, $\beta=0.00$ and $\gamma=-1.00$. This model has an effective nucleon mass of $0.7 M_N$, a compression modulus of $257.2 MeV$ (for comparison see refs. \cite{K1,K2,K3}) and a maximum neutron star mass of $1.48 M_\odot$ as shown in Fig. \ref{best}. This choice allow us to calibrate the model that can now be used, in a second step, to calculate many other neutron star properties such as radius, redshift, maximum rotational frequency, moment of inertia, cooling properties, among others.

At first sight the value of the maximum neutron star mass does not include the largest observed masses. But analyzing Fig. \ref{NSMass} one may reason that within the error bars the masses of most of the pulsars are compatible with $\sim1.4$ solar masses. Stars with larger masses may well contain a new kind of matter, such as quark matter, quark hadron mixtures or other exotic phases \cite{QGQHDStar1,QGQHDStar2,QGQHDStar3,QGQHDStar4}.

With the present work we established a model which is in agreement with nuclear phenomenology and observational data. Using the maximum mass as a selection criterion, determined by varying the model parameters $\lambda, \beta$ and $\gamma$, we were able to predict a compressibility of $K = 257,2 MeV$ from astrophysical arguments, in fairly good agreement with the limits given by other analysis.


\newpage

\begin{thebibliography}{99}
\bibitem{Yagi2006}
  K.~Yagi, T.~Hatsuda and Y.~Miake,
  Camb.\ Monogr.\ Part.\ Phys.\ Nucl.\ Phys.\ Cosmol.\  {\bf 23}, 1 (2005).

\bibitem{chiral1}
  P.~Papazoglou, S.~Schramm, J.~Schaffner-Bielich, H.~Stoecker and W.~Greiner,
  Phys.\ Rev.\  C {\bf 57}, 2576 (1998).

\bibitem{chiral2}
  P.~Papazoglou, D.~Zschiesche, S.~Schramm, J.~Schaffner-Bielich, H.~Stoecker and W.~Greiner,
  Phys.\ Rev.\  C {\bf 59}, 411 (1999).

\bibitem{han}
  M.~Hanauske, D.~Zschiesche, S.~Pal, S.~Schramm, H.~Stoecker and W.~Greiner,
  arXiv:astro-ph/9909052.

\bibitem{VADIWARA06}
  V.~A.~Dexheimer, C.~A.~Z.~Vasconcellos and B.~E.~J.~Bodmann,
  Int.\ J.\ Mod.\ Phys.\  D {\bf 16}, 269 (2007).

\bibitem{teseVAD2006} V.A. Dexheimer, {\sl Master Dissertation}, Instituto de F\'{\i}sica, Universidade Federal do Rio Grande do Sul, Porto Alegre, RS, Brazil (portuguese) (2006).


\bibitem{SU6}
  S.~Pal, M.~Hanauske, I.~Zakout, H.~Stoecker and W.~Greiner,
  Phys.\ Rev.\  C {\bf 60}, 015802 (1999)
  [arXiv:astro-ph/9905010].

\bibitem{Walecka} J.D. Walecka, {\sl Theoretical Nuclear And Subnuclear Physics} World Scientific Publishing Company; 2nd edition (2004).

\bibitem{ZM1}
  J.~Zimanyi and S.~A.~Moszkowski,
  Phys.\ Rev.\  C {\bf 42}, 1416 (1990).

\bibitem{ExpMod}
  A.~R.~Taurines, C.~A.~Z.~Vasconcellos, M.~Malheiro and M.~Chiapparini,
  Phys.\ Rev.\  C {\bf 63}, 065801 (2001)
  [arXiv:nucl-th/0010084].

\bibitem{Q0NS}
  C.~R.~Ghezzi,
  Phys.\ Rev.\  D {\bf 72}, 104017 (2005)
  [arXiv:gr-qc/0510106].

\bibitem{K1}
  D.~Vretenar, T.~Niksic and P.~Ring,
  Phys.\ Rev.\  C {\bf 68}, 024310 (2003)
  [arXiv:nucl-th/0302070].

\bibitem{K2}
  C.~Hartnack, H.~Oeschler and J.~Aichelin,
  Phys.\ Rev.\ Lett.\  {\bf 96}, 012302 (2006)
  [arXiv:nucl-th/0506087].

\bibitem{K3}
  P.~Arumugam, B.~K.~Sharma, P.~K.~Sahu, S.~K.~Patra, T.~Sil, M.~Centelles and X.~Vinas,
  Phys.\ Lett.\  B {\bf 601}, 51 (2004)
  [arXiv:nucl-th/0410029].

\bibitem{QGQHDStar1}
  G.~F.~Burgio,
  Nucl.\ Phys.\  A {\bf 749}, 337 (2005)
  [arXiv:nucl-th/0410040].

\bibitem{QGQHDStar2}
  A.~Drago, A.~Lavagno and G.~Pagliara,
  Nucl.\ Phys.\ Proc.\ Suppl.\  {\bf 138}, 522 (2005).

\bibitem{QGQHDStar3}
  J.~M.~Lattimer, M.~Prakash,
  Nucl.\ Phys.\ A {\bf 777}, 479 (2006).

\bibitem{QGQHDStar4}
  F.~Weber,
  Prog.\ Part.\ Nucl.\ Phys.\  {\bf 54}, 193 (2005)
  [arXiv:astro-ph/0407155].

\end{thebibliography}
\end{document}